# Design and Empirical Characterization of a Hardware-Realized Turing Machine with Automated Card-Based Programming

Agrima Regmi[1], Jenish Pant[2], Pratistha Sapkota[3] , Sanskriti Khatiwada[4], Binod Sapkota[5]

[1]*Dept. of Electronics and Computer Engineering, Thapathali Campus,* agrima.078bei003@tcioe.edu.np
[2]*Dept. of Electronics and Computer Engineering, Thapathali Campus,* jenish.078bei018@tcioe.edu.np
[3]*Dept. of Electronics and Computer Engineering, Thapathali Campus,* pratistha.078bei028@tcioe.edu.np
[4]*Dept. of Electronics and Computer Engineering, Thapathali Campus,* sanskriti.078bei037@tcioe.edu.np
[5]*Prof. Dept. of Electronics and Computer Engineering, Thapathali Campus,* bsapkota@tcioe.edu.np

***Abstract***

Physical implementations of Turing Machines remain rare, and existing electromechanical demonstrators and mechanical logic games typically require manual operator intervention, either to trigger each computational step or to reconfigure the state table, or both. This restricts prior physical models to short, operator-paced demonstrations and prevents autonomous execution of extended computations. This paper addresses that gap with a hardware Turing Machine that enables autonomous multi-step execution and reprogrammable optical input without manual intervention between programs. The system integrates an Arduino Mega for state-transition logic, dual NEMA 17 stepper motors for bidirectional tape actuation (shown experimentally to eliminate the positional drift seen in single-motor designs), infrared reflectance sensors for symbol detection, and an ESP32-CAM-based optical punched-card reader for automated state-table loading. Hole detection under non-uniform illumination used a Breadth-First Search flood-fill algorithm with local adaptive thresholding rather than fixed global thresholding, driven by the memory and library constraints of the ESP32-CAM's microcontroller environment; this improved card-decoding accuracy from 75% to 90% (100% with mechanical card flattening) on a 20-card test set. Mechanical evaluation showed fabrication accuracy of ±0.15 mm, rack-and-pinion positional error below 0.3 mm across 50 trials, and voltage supply stability within ±0.2 V under full system load. End-to-end computation was validated against a parallel software simulator (tlang), with all hardware outputs matching the simulated reference exactly across multiple test programs. The system advances prior physical Turing Machine demonstrations through autonomous execution, reprogrammable optical input, and quantitative evaluation of its mechanical, optical, and computational performance.

## 1. Introduction

The Universal Turing Machine (UTM), first introduced by Alan Turing in his groundbreaking 1936 paper *On Computable Numbers, with an Application to the Entscheidungsproblem [1]*, is a foundational concept in computer science that serves as a theoretical framework for understanding the boundaries of computation and the principles governing algorithmic processes. The UTM formalizes the notion of computation by using an abstract machine that consists of an infinite tape, segmented into discrete cells capable of holding symbols, and a movable head that reads, writes, and transitions between states based on a defined set of rules. These rules dictate the machine's behavior by associating the current symbol being read and the machine's state with specific actions, such as writing a symbol, moving the head, or changing states.

Formally, a Turing machine is defined as a 7-tuple [2]:

$$M = (Q,\Sigma,\Gamma,\delta,q0,\phi,F)$$

which specifies the machine’s states, symbols, transition rules, initial configuration, a blank symbol and accepting conditions.

Despite profoundly influencing the theoretical underpinnings of modern computing, the abstract representation of the machine as well as the absence of practical and interactive learning tools able to bridge the gap between theory and application posed significant challenges for students attempting to comprehend its operation and relevance. The absence of interactive, hardware-based tools that bridge theory and practice has been widely noted in computer science education.

* Corresponding author

While Turing Machines are commonly studied through software simulators, their physical implementation introduces a different set of requirements. A software simulator can execute a transition table directly, whereas a hardware implementation must translate each transition into coordinated physical actions such as moving the tape, detecting the current symbol, writing the next symbol, and changing state. Existing physical demonstrations have shown that these operations can be realized mechanically or electronically, but they often depend on manual intervention or simplified forms of program configuration. This limits their ability to demonstrate continuous, reprogrammable computation in a single physical system. The motivation of this work is therefore to develop a physical Turing Machine in which the transition table can be changed without rebuilding or manually reconfiguring the machine, while computation can proceed through multiple transitions without operator intervention.

This work makes several contributions to the physical realisation of programmable Turing Machine computation. On the hardware side, we present a complete UTM implementation using low-cost electromechanical components and demonstrate that a dual stepper motor configuration is necessary and sufficient to eliminate tape positional drift that single-motor designs cannot avoid. On the software and sensing side, we contribute a BFS-based flood fill decoder running on the ESP32-CAM for optical punched card reading, paired with local adaptive thresholding to handle real-world illumination variability. A custom Turing machine simulator (tlang) was developed in parallel as a pre-hardware verification environment, enabling transition rule validation before physical execution. Finally, we provide a systematic empirical characterisation of all subsystems, mechanical, optical, and electrical, offering a performance baseline for similar low-cost embedded hardware implementations.

## 2. Related Works

### *2.1 Early Theoretical Foundations*

The concept of the Turing Machine was introduced by Alan Turing in his seminal 1936 paper, which laid the foundation for computability theory. The term "Turing Machine" itself was coined by Alonzo Church in a review of Turing's work [3]. In the 1950s, Claude Shannon made significant contributions through his work on finite-state machines (FSMs), presented in Automata Studies [4], co-edited with John McCarthy. While Shannon's focus was primarily on theoretical frameworks, his research laid the foundation for hardware-based computational models. These insights were pivotal in bridging the gap between abstract automata and their physical implementations. Unlike these purely theoretical contributions, the present work targets physical realisation of the UTM model with empirically validated subsystem performance.

### *2.2 Software Simulators*

Numerous software tools have been developed to simulate Turing Machines. A notable example is JFLAP (Java Formal Languages and Automata Package), created by Sharon H. Rodger in the 1990s [5]. JFLAP enables users to experiment with finite automata, pushdown automata, multi-tape Turing Machines, and various grammars, and has been widely used in academia. More recently, tools such as FSM Builder [6] have extended this line of work by supporting autograded finite-automata exercises for classroom use, and a 2024 systematic review of automata theory education [7] confirms that the shortage of interactive, hands-on tools remains an open problem in the field. However, these efforts remain confined to the software domain. While JFLAP and its successors provide software-based simulation, the present work addresses the complementary challenge of physical execution, where mechanical tolerances, sensor noise, and power management introduce real-world constraints absent in software. To bridge this gap prior to hardware construction, a custom software simulator (tlang) was developed alongside the hardware, allowing digital validation of transition rules before physical deployment.

### *2.3 Physical Models*

Physical models of Turing Machines have been constructed to demonstrate Turing's concepts tangibly. Mike Davey's 2010 model [8] is a widely cited example that brings the abstract machine into physical form for educational purposes. However, Davey's implementation relies on mechanical components without electronic sensing or automated control, requiring manual intervention for state transitions. In contrast, the present work employs IR reflectance sensors and microcontroller-based logic to execute state transitions autonomously without operator input.

* *Corresponding author*

The Turing Tumble, a mechanical game designed to teach logic and programming principles, was demonstrated to be Turing-complete by Lenny Pitt in 2021 [9] when extended with an infinite board and unlimited pieces. However, the Turing Tumble requires manual ball placement for each computation step. The present implementation executes all steps autonomously through programmed logic, making it suitable for observing extended computations in real time.

A distinguishing feature of the present work, not found in prior physical UTM implementations, is the use of a camera-based optical reader (ESP32-CAM) for automated state table input via punched cards. This eliminates manual reprogramming between computation runs and provides a historically resonant input method that mirrors early computing systems.

### *2.4 Advanced Simulations*

With advancements in computational complexity theory, Turing Machine simulators have evolved to support applications in quantum computing, artificial intelligence, and cryptography. The Quantum Turing Machine (QTM), introduced by David Deutsch [10] and expanded by Bernstein and Vazirani [11], extends the traditional UTM model by incorporating quantum mechanical principles such as superposition and entanglement. Distributed extensions of the Turing Machine framework, such as Pineda's 2018 proposal [12], enhance applicability in associative memory and scalable parallel computation. The present work does not target these extensions but provides a physical baseline implementation that serves as a concrete reference for such theoretical comparisons.

## 3. Materials and Methods

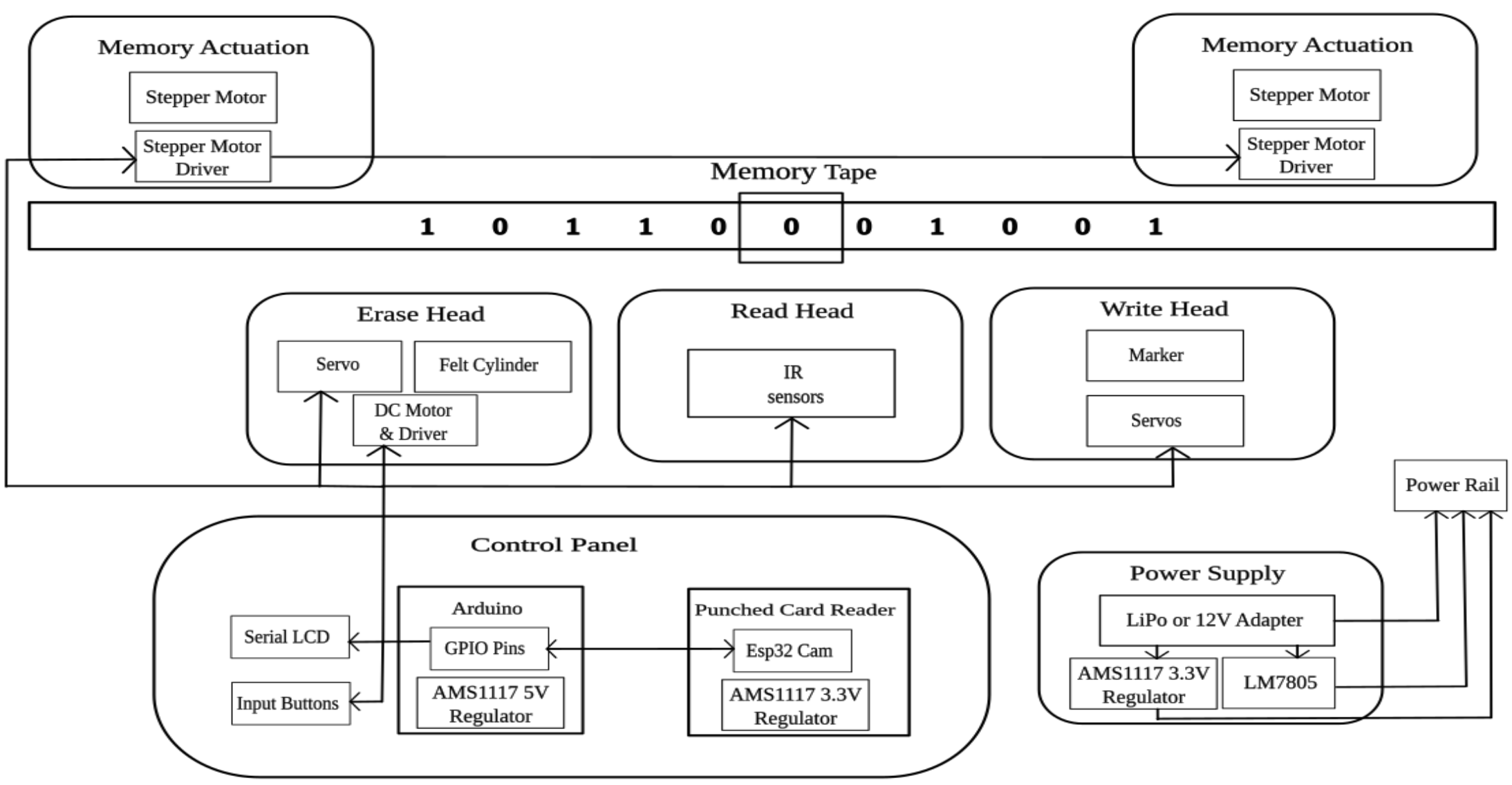

Figure 1: System Block Diagram

### *3.1 System Architecture*

The system architecture in Figure 1 was designed around four requirements: programmable transition-table input, reliable tape positioning, autonomous symbol read/write operations, and pre-hardware verification of computational behavior. These requirements are addressed through four interacting subsystems: an optical punched-card reader, a microcontroller-based control system, a motorized tape mechanism, and a software-based Turing Machine simulation and verification environment. The resulting workflow separates program acquisition and verification from physical execution while maintaining an interface between the digital transition table and the electromechanical machine.

#### *3.1.1 Memory Tape and Actuation System*

The memory tape serves as the primary data storage medium for the UTM. The tape is designed as a 35 mm laminated sheet divided into discrete cells, each representing a binary symbol (0 or 1), with an additional blank symbol included when required for computational purposes. A stepper motor, controlled by an Arduino through a dedicated motor driver, moves the tape left or right with high positional accuracy. The movement is regulated by a pulley system, where one full rotation of the stepper motor corresponds to the displacement of exactly one tape cell in either direction. This precise mechanical mapping ensures accurate alignment of each cell with the read, write, and erase heads [13]. As a result, the system enables reliable data access, consistent symbol manipulation, and repeatable tape positioning throughout the computation process.

* *Corresponding author*

#### *3.1.2 Read Head*

The read head employs two infrared (IR) reflectance sensors arranged linearly to detect the symbols encoded on the tape. Each tape cell is marked with predefined reflective patterns, and the sensors convert the reflected IR intensity into binary representations of the stored symbols. As the tape advances, the sensors sequentially sample each cell, enabling symbol identification based on reflectivity characteristics. The sensor output is mapped to tape symbols such that 0 is represented by Black–White, 1 is represented by Black–Black and B is represented by White–White (blank). The resulting binary data are transmitted to the control unit, where they are decoded and used for state transition and computational processing.

#### *3.1.3 Write Head*

The write head is equipped with a marker controlled by two servo motors, which guide the marker along two axes using a dual rack and pinion system. These servo motors precisely position the marker to either write a new symbol or leave the existing symbol unchanged, depending on the instructions received from the control unit. This controlled movement allows accurate symbol placement within each tape cell and enables the machine to reliably modify its contents as specified by the defined transition rules of the UTM.

#### *3.1.4 Erase Head*

The erase head consists of a felt cylinder driven by a DC motor and actuated along one axis using a rack and pinion mechanism. This felt cylinder is specifically designed to make uniform contact with the tape surface, ensuring effective symbol removal without damaging the tape. Upon receiving an erase command from the control unit, the cylinder moves into position, engages with the tape, and clears the symbol in the current cell, thereby preparing it for accurate and clean overwriting in subsequent operations.

#### *3.1.5 Control Panel*

An Arduino micro-controller functions as the central control unit of the system, managing the operation of all components based on the transition rules and the initial memory configuration, which is derived from the punched card input. The control panel features input buttons that enable the user to start, stop, step through the program, load the state table, or reset the machine. An LCD display provides real-time feedback to the user, displaying the current state of the Turing machine, including the content of the tape, the control state, and the operational status of the machine.

#### *3.1.6 Mechanical Design*

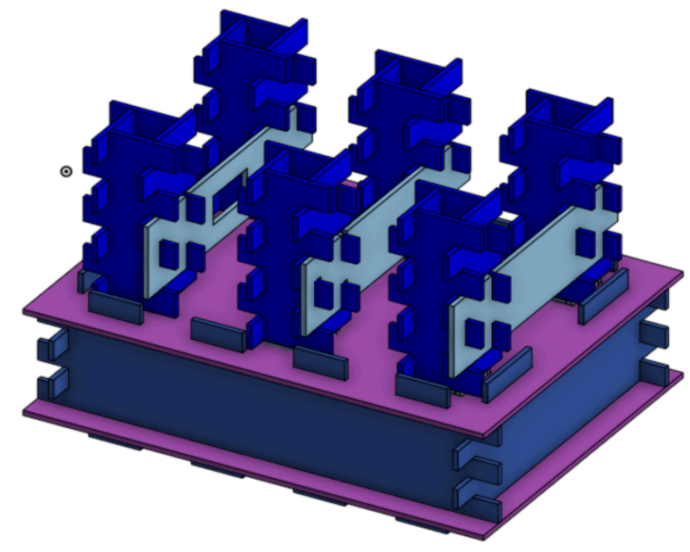

(a) Isometric View

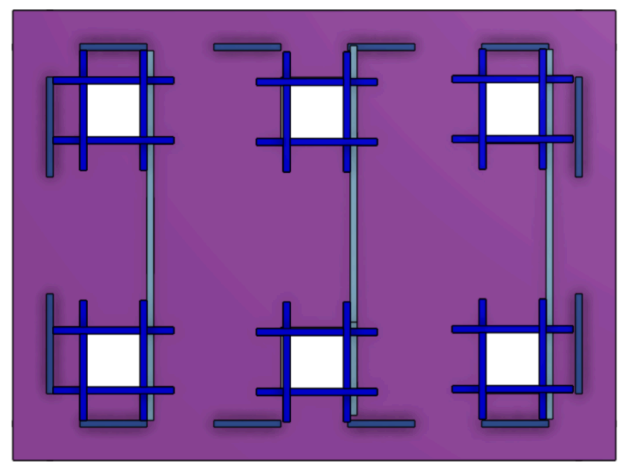

(b) Top View of the Base

Figure 2: CAD model of the machine base designed in Onshape

The mechanical structure of the proposed system was designed using Onshape CAD to ensure dimensional accuracy and reliable integration of all components. The base structure has overall dimensions of 180 mm × 135.1mm × 50mm, providing adequate support for the transport mechanism, sensors, and control electronics. Figure 2 presents the isometric and orthographic views of the developed model.

* Corresponding author

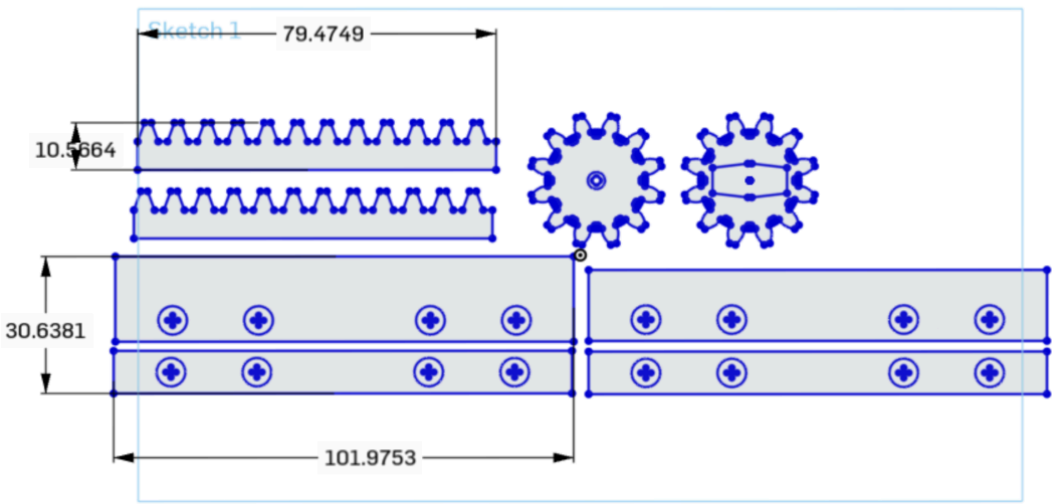

Figure 3: CAD model of rack and pinion mechanism for linear motion conversion

The card transport mechanism requires controlled linear motion for accurate positioning. To achieve this, a rack-and-pinion mechanism driven by servo motors was employed. The mechanism converts the rotational motion of the servo shaft into linear displacement, enabling precise movement of the transport assembly. The detailed design of the rack-and-pinion system is shown in Figure 3.

### *3.2 Punched Card Reader*

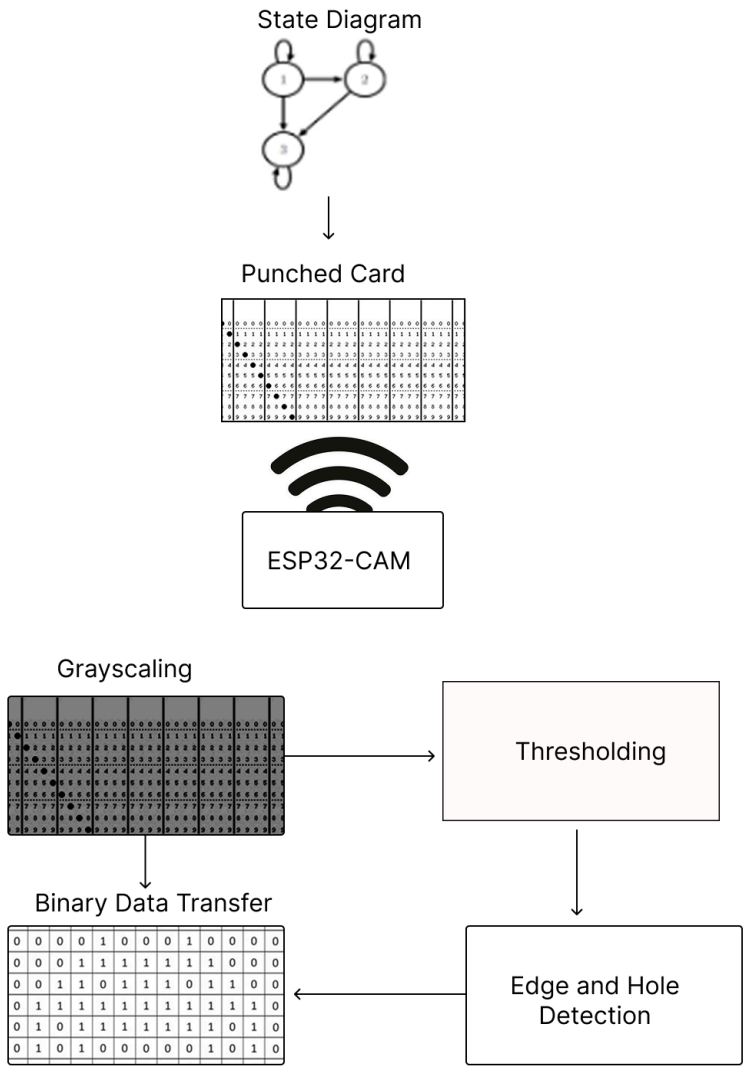

Figure 4: Working flow of Punched Card Reader

The punched card reader interprets binary data encoded on punched cards, the UTM's primary input medium. Punched cards were historically used as programmable input in early computing systems prior to electronic memory [14]; inspired by this approach, the system uses an ESP32-CAM module [15] to capture high-resolution images of each card and digitize its hole pattern, detecting the presence or absence of holes that correspond to specific binary values.

Captured images undergo grayscale conversion, noise reduction, and thresholding to distinguish holes from the background. The resulting hole pattern is translated into a binary sequence and decoded into the corresponding transition rules.

Decoded rules are transmitted to the Arduino-based control unit over a serial interface, which uses them to manage state transitions, tape movement, and symbol manipulation. This automated pipeline removes the need for manual input, improving reliability, accuracy, and efficiency of instruction loading.

* *Corresponding author*

### 3.3 Image Processing Pipeline

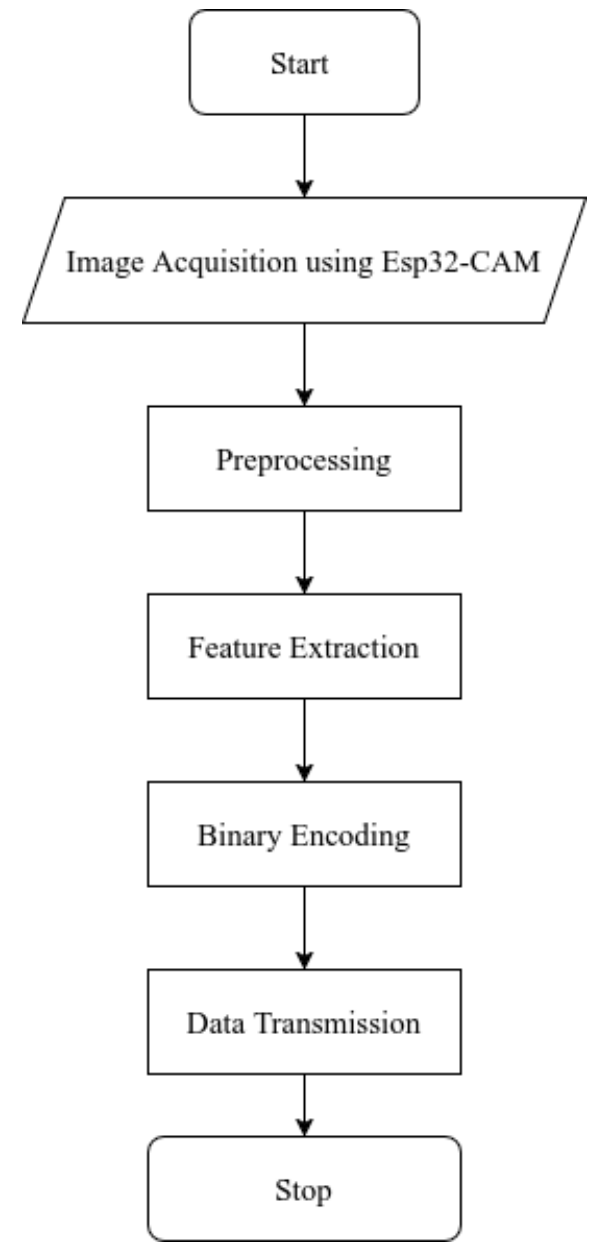

Figure 5: Image Processing Flowchart

#### 3.3.1 Grayscale Conversion

The RGB image is converted to a single-channel grayscale image using the luminosity-weighted method, which accounts for the human eye's differential sensitivity to colour channels and, in practice, maximises contrast between black ink and white card substrate [16]:

$$Grayscale = 0.299R + 0.587G + 0.114B \quad \text{(Equation 1)}$$

#### 3.3.2 Binarization

Two binarization approaches were evaluated. Global thresholding applies a single fixed intensity threshold (T = 128) across the entire image [17]. While computationally simple, this approach was found to produce false positives and misses under non-uniform card illumination, particularly near card edges where reflectance varies. Local (adaptive) thresholding divides the image into sub-regions and computes an independent threshold for each region based on local pixel intensity statistics [18]. This method proved significantly more robust under the variable lighting conditions of the deployment environment and was adopted as the primary binarization method [19].

#### 3.3.3 Flood Fill Hole Detection

After binarization, punched holes appear as connected regions of black pixels. Detecting these regions is a connected-component labeling problem, in which pixels belonging to the same object are grouped based on spatial connectivity [20]. The challenge is to reliably identify and size these regions while rejecting noise artefacts. A Breadth-First Search (BFS) flood fill algorithm was implemented for this purpose. Hole detection required a connectivity-based region labeling approach. OpenCV's connected-component and contour functions were unavailable on the ESP32-CAM's microcontroller environment, ruling out the standard approach. A two-pass labeling algorithm was considered but required an additional equivalence-resolution step; BFS flood fill was adopted instead, as it produced equivalent labeling in a single pass with O(W×H), where W and H are the image width and height, additional memory per pixel (a visited flag) and a queue bounded by the largest hole's pixel area, which is small relative to frame size, which suited the platform's limited RAM. DFS was rejected on the same grounds: its stack depth scales with the hole area, making worst-case memory use less predictable than BFS's queue-based traversal. The algorithm operates as follows: beginning from the first unvisited black pixel (the seed point), a queue is initialised for BFS traversal. A hole-size counter (holeSize) is set to zero. For each dequeued pixel, its four connected neighbours (above, below, left, right) are inspected. Neighbours with pixel value 0 (black, unvisited) are enqueued and immediately marked as visited by setting their value to 255, preventing revisitation. holeSize is incremented for each pixel processed. Traversal continues until the queue is empty, at which point holeSize represents the total area of the detected connected region. The process repeats for all remaining unvisited black pixels. This process can be formalised as:

$$FloodFill(x, y) = v, \text{ if } I(x, y) = I(x_0, y_0)$$

$$FloodFill(x', y') \text{ for all 4-connected neighbours } (x', y') \text{ of } (x, y)$$

where I is the binarized image, $(x_0, y_0)$ is the seed pixel, and v is the fill value (255). Connected regions with holeSize below a minimum pixel-area threshold are discarded as noise. Regions above the threshold are classified as valid punched holes. This approach correctly handles irregularly shaped holes and is robust to minor ink bleeding around hole boundaries.

* Corresponding author

#### *3.3.4 Binary Encoding and Transmission*

Detected hole positions are mapped to their row and column indices within the card grid. Each cell position encodes one field of a transition rule (current state, read symbol, write symbol, head direction, next state) according to a predefined column assignment scheme. Hole present = binary 1; hole absent = binary 0. The assembled binary string for each transition rule is serialised and transmitted to the Arduino Mega over UART. The Arduino's Serial.read() function receives the data, deserialises the transition table, and loads it into RAM for execution.

### *3.4 Turing Machine Simulation - tlang*

To complement the hardware implementation, a software-based Turing Machine simulator with an associated custom language was developed to observe and verify machine behavior digitally [21]. The simulator is used to validate transition rules, test algorithms, and establish a reference for the expected hardware operation. It was implemented in C for its efficiency, simplicity, and suitability for potential porting to embedded hardware during development.

#### *3.4.1 Design and Features*

The simulator replicates the fundamental components of a Turing Machine, including an extendable memory tape, a movable head, and a state transition mechanism defined through a compact domain-specific language. This language specifies symbol modification, head movement, and next-state transitions for each state–symbol pair. The instruction set is minimized to the keywords {F, N, R, L, S, H}, enabling concise representation of arbitrary state tables. The simulator executes operations step-by-step, allowing detailed observation of state transitions and tape modifications.

#### *3.4.2 Application and Insights*

The simulator was extensively used to test and validate transition rules prior to hardware implementation, helping identify logical errors and edge cases. This process ensured algorithm correctness and guided key design decisions, thereby improving the reliability and feasibility of the physical Universal Turing Machine.

#### *3.4.3 Visualization*

The simulator features a graphical interface that displays :

- The current tape content, with the active cell highlighted.
- The current state of the Turing Machine.
- The head's position on the tape.

* *Corresponding author*

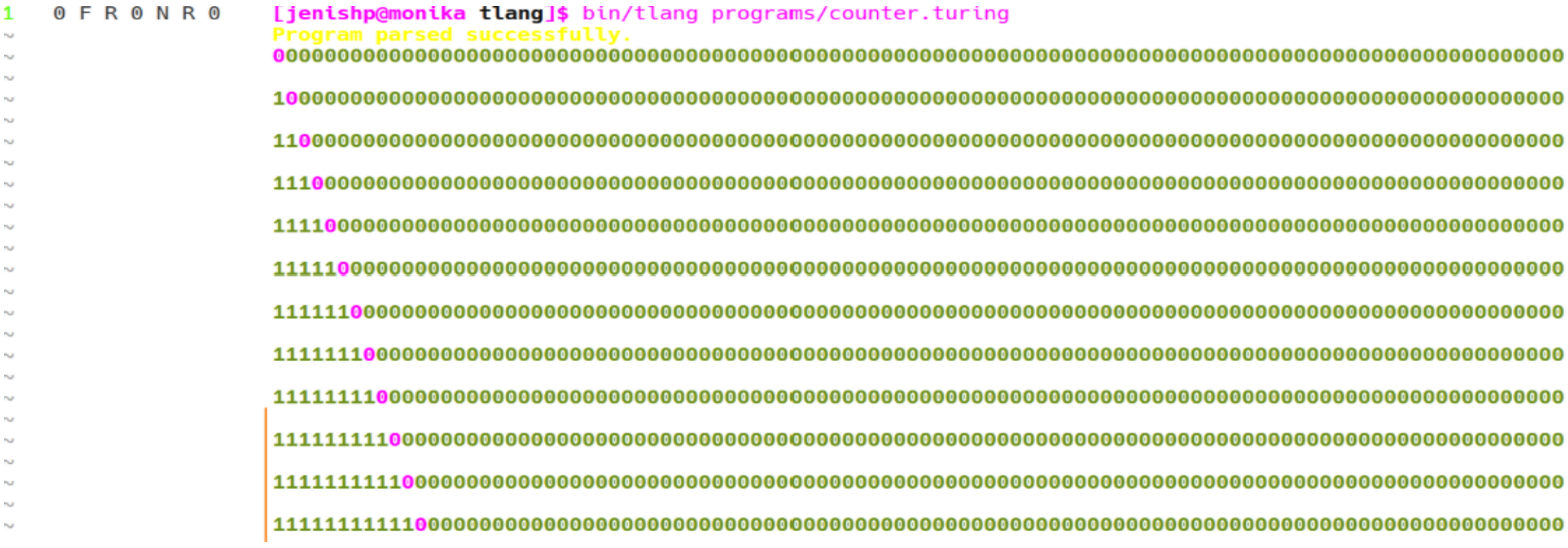

Figure 6: tlang simulator UI

## 4. Results

Each subsystem was independently characterised before full-system integration. Results are reported below with the test methodology used for each measurement.

### *4.1 Structural Base*

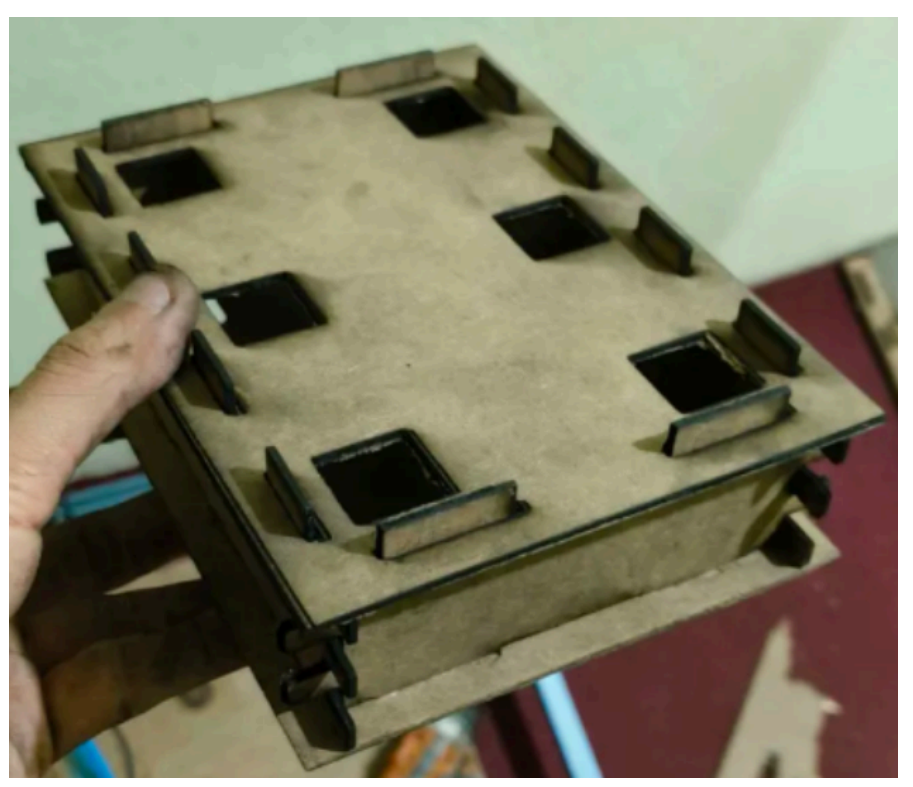

Figure 7: Structural Base

The system base was fabricated from Medium-Density Fiberboard (MDF) using a $CO_2$ laser cutter operating at 50 W and 300 mm/min feed rate, producing a kerf width of approximately 1 mm, which was compensated for in the CAD design. Post-fabrication dimensional measurements at 50 points across the base confirmed an average deviation of ±0.15 mm from the CAD nominal dimensions, validating the laser-cutting process parameters. Slot positions were verified by assembling all mechanical components and checking clearance fit. Structural load testing demonstrated stability under applied loads up to 2 kg, with minor flexing onset above 3 kg, well above the operational load of the assembled machine (approximately 0.8 kg). Minimal edge charring was observed, confirming that laser parameters were within acceptable limits.

* *Corresponding author*

### *4.2 Linear Movement Mechanism*

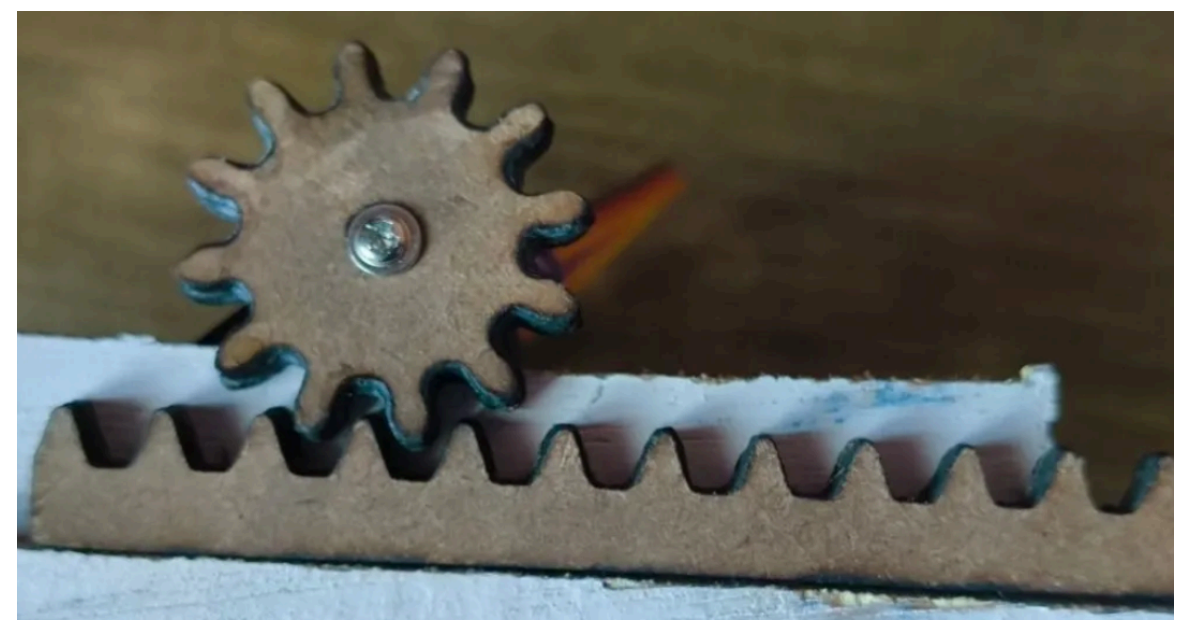

Figure 8: Linear Movement Mechanism for Tape Modification

The rack-and-pinion write-head actuation mechanism was fabricated from 2 mm MDF. Positional accuracy was evaluated over 50 controlled servo-driven displacements, measuring the deviation between commanded and actual head position using calipers. The mean positional error was 0.28 mm (below the 0.3 mm target), with a maximum observed error of 0.41 mm. Response time was measured by commanding 10 mm displacements repeatedly and averaging over 30 trials, yielding a mean response time of 150 ms. Speed control via PWM duty cycle exhibited a near-linear velocity response ($R^2 \approx 0.97$), with minimal overshoot at all tested duty cycles. Load-bearing tests showed smooth operation up to 50 g applied load, minor stress at 100 g, and an upper operational limit of approximately 200 g, above which rack deflection became measurable. Lubrication of the pinion gear and addition of a lateral guide rail improved long-term repeatability.

### *4.3 Electrical System*

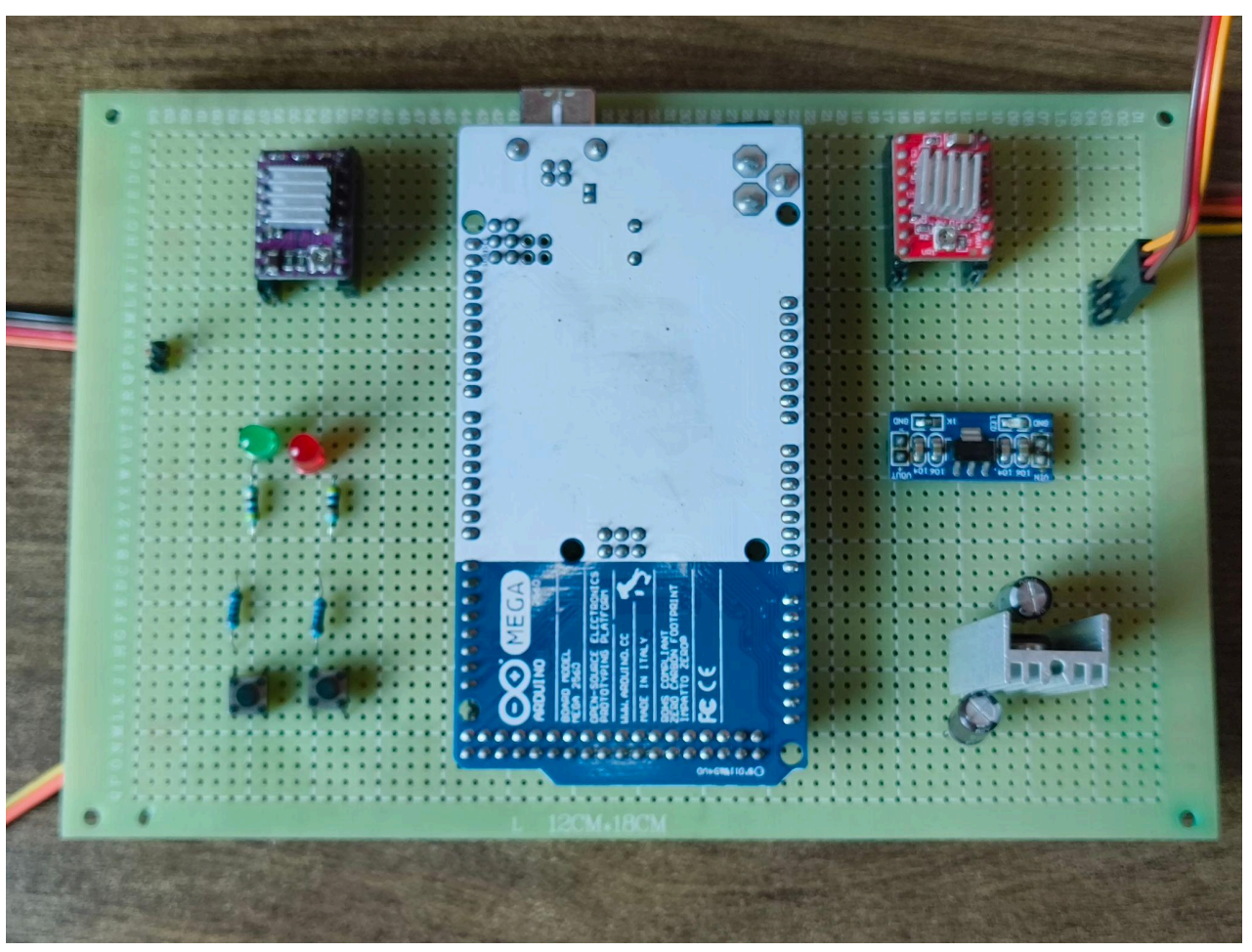

Figure 9: Main Control Board of the System

The control board integrates the Arduino Mega, ESP32-CAM, two A4988 stepper drivers, three 9g servo motors, IR sensors, and power regulation circuits on a matrix perfboard. Power is distributed at three voltage rails: 12 V for stepper motors, 5 V (via LM7805 regulators) for logic and servos, and 3.3 V (via AMS1117) for the ESP32-CAM. Under full-load conditions with all actuators simultaneously active, voltage measurements at each rail showed deviations within ±0.2 V of nominal, confirming adequate regulation. Stepper driver peak current was set to 1.8 A via the A4988 VREF trimpot, preventing overheating without sacrificing torque. Heat sinks were mounted on both A4988 drivers. Thermal imaging during 30-minute continuous operation confirmed all components remained within safe operating temperature limits. Decoupling capacitors (100 µF) placed across motor power rails mitigated transient voltage spikes that had caused ESP32-CAM resets during initial testing.

* Corresponding author

### *4.4 Memory Tape*

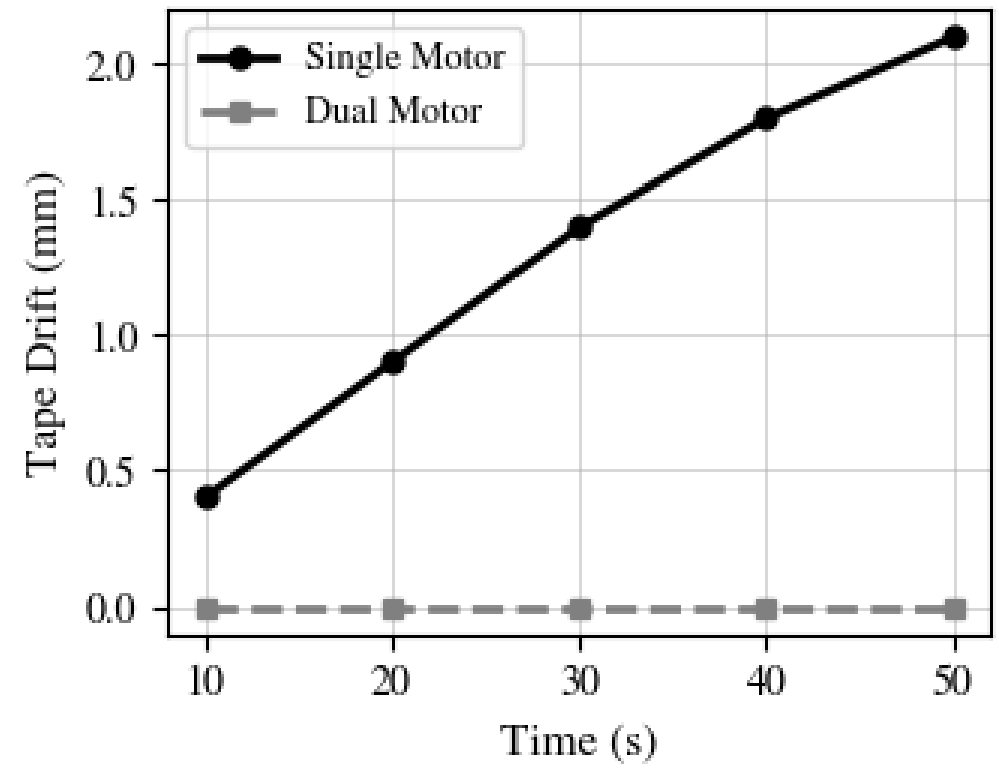

Figure 10: Comparison of tape drift between single-motor and dual-motor tape transport systems

The lamination sheet tape was tested for positional consistency over 200 consecutive cell-advance cycles. No measurable slippage was detected when the dual-motor configuration was active. As shown in Figure 10, with the original single-motor design, cumulative positional drift of up to 2.1 mm was observed over 50 cycles, causing the read head to straddle cell boundaries and produce symbol misreads. The dual-motor fix eliminated this drift entirely [22]. Dry-erase marker symbols applied by the write head were consistently legible to the IR sensors and fully removable by the felt erase head across at least 50 write-erase cycles per cell with no tape surface degradation.

### *4.5 Read/Write Head*

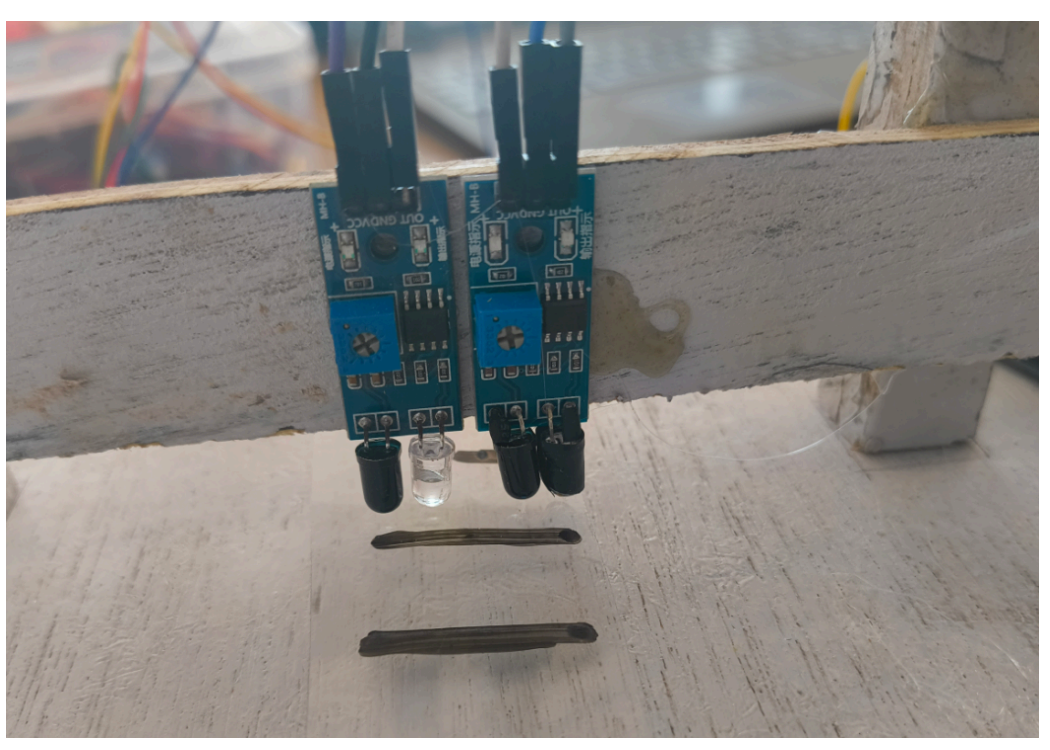

Figure 11: Dual IR alignment for tape symbol detection

Figure 11 shows the dual IR sensors alignment for detecting symbols on the memory tape. IR sensor performance was characterised by writing known symbol sequences on the tape and measuring the read accuracy under controlled and variable lighting. Under diffuse indoor lighting with the sensor enclosure fitted,symbol classification accuracy was 100% across 150 consecutive reads (n=150;95%CI: 97.5%–100%) (50 per symbol type: 0, 1, Blank). Under direct overhead fluorescent lighting without enclosure shielding, misread rates of approximately 8% were observed, primarily in the Black–White (symbol 0) case where ambient light partially masked the white-region reflection difference. Fitting the partial enclosure reduced this to below 1%. The servo-controlled write mechanism produced legible, within-boundary symbols in 98% of write operations across 100 trials (n=100; 95% CI: 93.0%–99.4%) ; the 2% failure rate was attributed to servo position drift under extended continuous operation, addressable by periodic recalibration.

* *Corresponding author*

#### *4.6 Punched Card Reader*

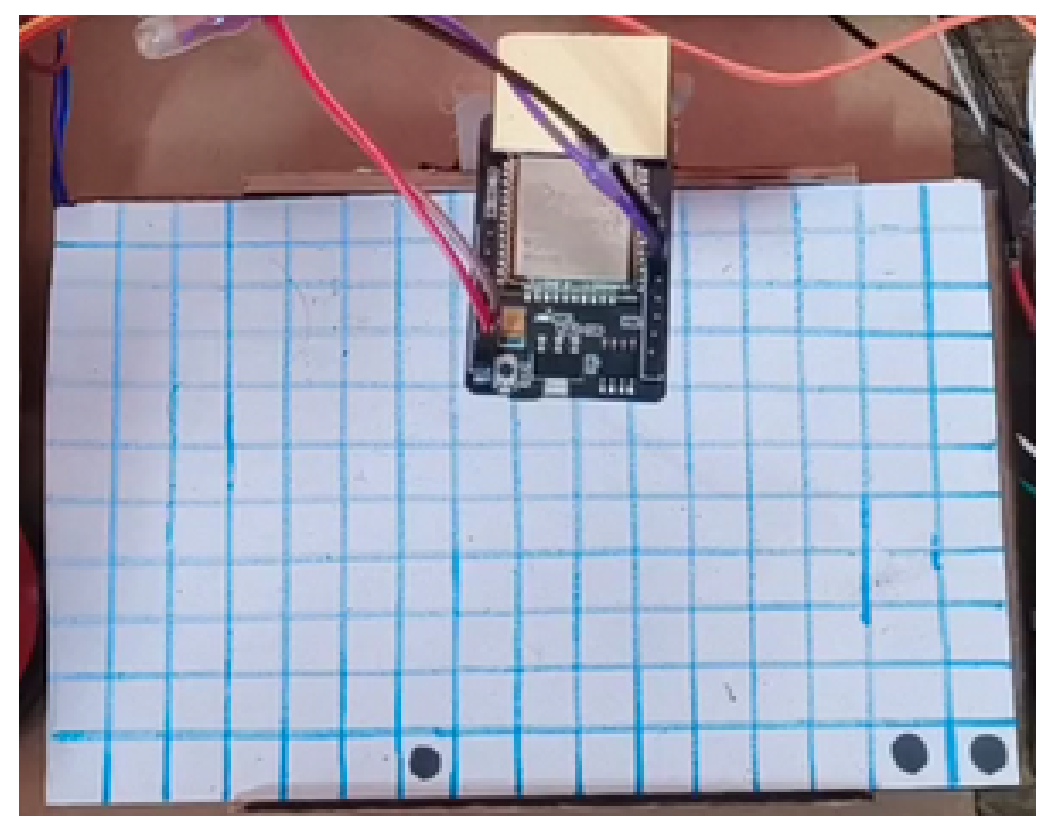

Figure 12: ESP32-CAM image acquisition setup used for punch card digitization

Figure 12 shows the ESP32-CAM-based image acquisition setup used for punch card digitization. During operation, the punch card is positioned beneath the camera, which captures a top-down image of the card containing the punched-hole pattern. The captured image is then processed through grayscale conversion, adaptive thresholding, and BFS flood-fill hole detection to extract the card's transition table for subsequent finite state machine reconstruction. Card reading accuracy was evaluated using a test set of 20 cards with known transition tables (4–6 states each, 15–25 transitions per card). The BFS flood fill decoder correctly identified all holes in 18 of 20 cards (90%; 95% CI: 69.9% – 97.2%) , with two cards producing single-bit errors attributable to shadow artefacts from card edge curl under the fixed illumination. When cards were held flat (manual pressure or card clamp), accuracy rose to 100% across all 20 cards. Global thresholding alone achieved 75% card accuracy on the same test set, confirming the superiority of local adaptive thresholding under real conditions. The full card decoding and UART transmission cycle completed in approximately 1.2 seconds per card.

#### *4.7 End-to-End Computation*

To validate full system operation, a binary increment algorithm was implemented as the primary test case. The state table (4 states, 8 transitions) was encoded on a punched card, loaded via the ESP32-CAM reader, and executed on a tape initialised with the binary value 0110 (decimal 6). The machine executed 12 state transitions over approximately 38 seconds. This corresponds to an average execution time of approximately 3.17 seconds per transition, including symbol reading, tape movement, write/erase operations, and state updates, producing the output 0111 (decimal 7) on the tape, consistent with the expected result validated in tlang simulation. All state transitions matched the tlang reference exactly. A second test case, unary addition of two operands separated by a blank, was executed successfully for operand pairs up to 5+4. These demonstrations confirm that the system executes algorithms correctly end-to-end and that the tlang simulator serves as a reliable pre-hardware verification tool.

#### *4.8 System Performance Summary*

Table 1. Summary of measured performance of the major subsystems comprising the hardware Universal Turing Machine.

| Subsystem | Metric | Result |
|---|---|---|
| Structural Base | Dimensional Accuracy | ±0.15 mm |
| Linear Movement Mechanism | Mean Positional Error | 0.28 mm |
| Linear Movement Mechanism | Maximum Error | 0.41 mm |
| Electrical System | Voltage Stability | ±0.2 V |
| Memory Tape | Positional Drift | Not detected |

* *Corresponding author*

| Read Head | Symbol Classification Accuracy | 100% (150 reads) |
| --- | --- | --- |
| Write Head | Successful Write Operations | 98% (100 writes) |
| Punched Card Reader | Card Decoding Accuracy(20 cards) | 90% (100% with card flattening) |
| Punched Card Reader | Decode Time | 1.2 s/card |

The results demonstrate that all major subsystems met their design objectives. The highest reliability was observed in the read-head subsystem, which achieved perfect symbol classification under controlled operating conditions. The punched card reader achieved the lowest overall accuracy due to occasional shadow artefacts caused by card edge curling; however, this issue was eliminated when the card was mechanically constrained during scanning.

### *4.9 Simulator–Hardware Consistency*

To verify that the hardware implementation correctly executed the intended transition rules, all test programs were first validated using the tlang simulator and subsequently executed on the physical Universal Turing Machine.

| Test Case | Initial Tape | Expected Output(tlang) | Hardware Output |
| --- | --- | --- | --- |
| Binary Increment | 0110 | 0111 | 0111 |
| Unary Addition( 5 + 4) | 11111B1111 | 111111111 | 111111111 |

In all evaluated test cases, the final tape contents produced by the hardware implementation matched the outputs generated by the simulator. No discrepancies were observed between the simulated state transitions and the transitions executed by the physical system. These results confirm that the Arduino-based control logic correctly implements the transition function validated within the tlang environment and demonstrate the effectiveness of simulator-assisted hardware verification.

### *4.10 Reliability Assessment*

To evaluate operational reliability, repeated subsystem tests were conducted under normal operating conditions. The memory tape was advanced through 200 consecutive cell movements while monitoring positional alignment. No measurable positional drift was observed in the dual-motor configuration. Similarly, the write–erase subsystem completed more than 50 write–erase cycles per cell without visible tape degradation.

The punched card reader successfully decoded 18 of 20 test cards under unconstrained conditions, corresponding to an accuracy of 90%. Analysis of the two decoding failures revealed that both errors were caused by card edge curling, which introduced shadows during image acquisition. When the cards were manually flattened, all 20 cards were decoded correctly, resulting in 100% decoding accuracy. These results indicate that the primary limitations of the current implementation arise from mechanical and environmental factors rather than computational or algorithmic deficiencies.

## 5. Discussion

### *5.1 Comparison with Prior Physical Implementations*

Compared to Mike Davey's physical Turing Machine model, the present implementation adds three significant capabilities: (1) automated symbol reading via electronic IR sensors rather than visual or manual inspection; (2) automated state-transition execution via microcontroller logic without operator intervention between steps; and (3) a reprogrammable input system via optical punched card reading, enabling different algorithms to be loaded without hardware modification. These additions make the machine suitable for observing extended, multi-step computations, whereas prior simple models were typically demonstrated for only a small number of manually advanced steps.

* *Corresponding author*

The Turing Tumble, while an engaging educational tool, requires a human operator to place components and trigger each computation step. The present machine operates autonomously after state table loading, which is more representative of how a real computational device functions and makes it more useful for demonstrating the concept of algorithmic execution to students.

Table 2. Comparison with Existing Physical Turing Machine Implementations

| Feature | Mike Davey(2010) | Turing Tumble | Present Work |
|---|---|---|---|
| Autonomous Execution | No | No | Yes |
| Electronic Symbol Detection | No | No | Yes |
| Microcontroller-Based Control | No | No | Yes |
| Reprogrammable State Tables | No | No | Yes |
| Optical Input System | No | No | Yes |
| Simulator-Assisted Verification | No | No | Yes |
| Quantitative Evaluation | Limited | No | Yes |

As shown in Table 2, the proposed system extends previous physical implementations by incorporating autonomous operation, electronic sensing, programmable state-table loading, and quantitative subsystem evaluation. These additions allow the machine to execute longer computations without operator intervention while providing a reproducible platform for studying computational theory in a physical environment.

***5.2 Engineering Rationale and Design Trade-offs***

The single-motor design drove the tape from one end, allowing slack to accumulate at the free end under acceleration; driving both ends under tension removes this slack entirely, explaining the complete elimination of drift in Figure 10. The gain from adaptive over global thresholding (75%→90%) reflects the fixed onboard illumination's spatial intensity gradient, which a per-region adaptive threshold compensates for but a single global threshold cannot; residual global-threshold errors were concentrated at card edges, where the gradient was steepest. BFS flood fill's robustness follows from relying on pixel adjacency rather than assumed hole geometry, so irregular boundaries from ink bleed or laser-cut imprecision are absorbed without affecting detection.

Three trade-offs shaped the design. 2 mm MDF minimised fabrication cost but limited load tolerance (~200 g) and long-term durability; a rigid material such as acrylic would trade cost for lifespan. Fixed onboard illumination kept the card reader compact but left curl-induced shadowing as the residual error source (Section 4.6). Finally, the ~1 symbol/s execution rate is a direct consequence of mechanically actuated read/write/erase heads rather than electronic memory, an intrinsic cost of physically instantiating Turing-machine operations rather than a design deficiency.

***5.3 Hardware vs. Simulator Consistency***

All transition rules used in hardware tests were first validated in tlang. In hardware runs, all deviations from the expected output were traced to physical subsystem faults (sensor misread under poor lighting, tape slippage in early single-motor trials), not to logical errors in the state tables. This validates the design methodology of using a software simulator as a pre-hardware verification stage and confirms that the hardware correctly implements the theoretical transition function when subsystems operate within their characterised tolerances.

* *Corresponding author*

### *5.4 Limitations*

Several limitations remain. The finite tape length restricts the machine to computations solvable within the available number of cells. The current punched card format supports a maximum of 22 states, limiting algorithm complexity. IR sensor performance degrades under direct high-intensity lighting, requiring either controlled ambient conditions or improved enclosure design. The mechanical write system operates at approximately one symbol per second, making the machine impractical for any computation beyond educational demonstration. The rack-and-pinion mechanism fabricated from 2 mm MDF showed deflection above 200 g load and long-term wear under sustained use; a more durable material such as acrylic or aluminium extrusion would improve longevity.

## 6. Conclusion

This paper presented the design, implementation, and empirical evaluation of a complete hardware Universal Turing Machine. The system integrates dual NEMA 17 stepper motors for precise tape actuation, IR reflectance sensors for symbol reading, servo-actuated write and erase heads, and an ESP32-CAM optical punched card reader with a BFS flood fill decoding pipeline. Measured fabrication accuracy of ±0.15 mm, rack-and-pinion positional error below 0.3 mm, and voltage supply stability within ±0.2 V demonstrate robust subsystem performance. End-to-end computation was verified for a binary increment algorithm (12 transitions, output matching tlang simulation) and a unary addition algorithm. The dual stepper motor design, developed in response to single-motor tape drift, and the local adaptive thresholding pipeline, developed to overcome illumination non-uniformity in card reading, represent two engineering contributions with direct applicability to similar low-cost embedded sensing systems. The accompanying tlang simulator provided pre-hardware verification and confirmed consistency between theoretical transition rules and physical execution in all tested cases. Experimental evaluation demonstrated reliable subsystem operation across mechanical, electrical, sensing, and computational components. Furthermore, complete agreement between simulator outputs and hardware execution confirms the correctness of the implementation and validates the simulator-assisted design methodology adopted throughout development. The system serves as a functional, low-cost, and educationally valuable physical realisation of the Universal Turing Machine. Future work will focus on increasing tape capacity, improving card-reader robustness under uncontrolled lighting conditions, and replacing MDF-based mechanical components with more durable materials to support long-term operation.

* Corresponding author

* *Corresponding author*